
%
%
%
%
\input harvmac
%
%
\def\listrefs{\footatend\bigskip\bigskip\immediate\closeout\rfile\writestoppt
\baselineskip=12pt\parskip=2pt plus 1pt
\noindent{{\bf References}}\medskip{\frenchspacing%
\parindent=20pt\escapechar=` \input \jobname.refs\vfill\eject}\nonfrenchspacing}
%
%
%
\def\RF#1#2{\if*#1\ref#1{#2.}\else#1\fi}
\def\NRF#1#2{\if*#1\nref#1{#2.}\fi}
\def\refdef#1#2#3{\def#1{*}\def#2{#3}}
%
%
\def\ts{\hskip .16667em\relax}
\def\AJM{{\it Am.\ts J.\ts Math.\ts}}
\def\ASPM{{\it Advanced Studies in Pure Mathematics\ts}}
\def\CMP{{\it Comm.\ts Math.\ts Phys.\ts}}

\def\IJMP{{\it Int.\ts J.\ts Mod.\ts Phys.\ts}}
\def\JP{{\it J.\ts Phys.\ts}}
\def\NP{{\it Nucl.\ts Phys.\ts}}
\def\PL{{\it Phys.\ts Lett.\ts}}
\def\PJM{{\it Pacific\ts J.\ts Math.\ts}}
\def\PNAS{{\it Proc.\ts Natl.\ts Acad.\ts Sci.\ts USA\ts}}

\def\PSPM{{\it Proc.\ts Symp.\ts Pure\ts Math.\ts}}
\def\TAMS{{\it Trans.\ts Amer.\ts Math.\ts Soc.\ts}}
\def\Zm{Zamolodchikov}
\def\AZm{A.\ts B.\ts \Zm}

\def\me{P.\ts E.\ts Dorey}
\def\dur{H.\ts W.\ts Braden, E.\ts Corrigan, \me\ and R.\ts Sasaki}
%
%
\refdef\rBa\Ba{H.\ts W.\ts Braden, \JP {\bf A25} (1992) L15}
\refdef\rBc\Bc{N.\ts Bourbaki, {\it Groupes et alg\`ebres de Lie} {\bf
 IV, V, VI,} (Hermann, Paris 1968)}
\refdef\rBCa\BCa{R. Brunskill and A. Clifton-Taylor, {\it English Brickwork}
 (Hyperion 1977)}
\refdef\rBCDSc\BCDSc{\dur, \NP {\bf B338} (1990) 689}
\refdef\rCf\Cf{R.\ts Carter, {\it Simple Groups of Lie Type}, (Wiley
 1972)}
\refdef\rCj\Cj{J.\ts L.\ts Cardy, \NP {\bf B240} (1984) 514}
\refdef\rCk\Ck{J.\ts L.\ts Cardy, \NP {\bf B270} (1986) 186}
\refdef\rCl\Cl{J.\ts L.\ts Cardy, \NP {\bf B275} (1986) 200}
\refdef\rCm\Cm{J.\ts L.\ts Cardy, \NP {\bf B324} (1989) 581}
\refdef\rCDa\CDa{E.\ts Corrigan and \me, \PL {\bf B273} (1991) 237}
\refdef\rCIZa\CIZa{A. Cappelli, C. Itzykson and J-B Zuber, \CMP
 {\bf 113} (1987) 1}
\refdef\rDc\Dc{\me, \NP {\bf B358} (1991) 654}
\refdef\rDd\Dd{\me, \NP {\bf B374} (1992) 741}
\refdef\rDe\De{P.\ts Di\ts Francesco, \IJMP {\bf A7} (1992) 407}
\refdef\rDZa\DZa{P.\ts Di\ts Francesco and J.-B.\ts Zuber, \NP {\bf B338}
(1990) 602}
\refdef\rDZb\DZb{P.\ts Di\ts Francesco and J.-B.\ts Zuber,
 `SU(N) lattice integrable models and modular invariance',
 Proceedings of the Trieste conference on recent developments in
 conformal field theories, October 1989}
\refdef\rFb\Fb{M.\ts D.\ts Freeman, \PL {\bf B261} (1991) 57}
\refdef\rFLOa\FLOa{A.\ts Fring, H.\ts C.\ts Liao and D.\ts I.\ts Olive,
   \PL {\bf B266} (1991) 82}
\refdef\rFOa\FOa{A.\ts Fring and D.\ts I.\ts Olive,
 `The fusing rule and the scattering matrix of affine Toda theory',
 preprint Imperial/TP-91-92/08}
\refdef\rIa\Ia{C.\ts Itzykson, \ASPM {\bf 19} (1989) 287}
\refdef\rKb\Kb{B.\ts Kostant, \AJM {\bf 81} (1959) 973}
\refdef\rKc\Kc{B.\ts Kostant, \PNAS {\bf 81} (1984) 5275}
\refdef\rKd\Kd{B.\ts Kostant, {\it Ast\'erisque, hors s\'erie,}
  1985, 209 - 255}
\refdef\rLWb\LWb{W.\ts Lerche and N.\ts P.\ts Warner, \NP {\bf B358}
(1991) 571}
\refdef\rMd\Md{N.\ts J.\ts MacKay, \NP {\bf B356} (1991) 729}
\refdef\rMj\Mj{J.\ts McKay, \PSPM {\bf 37} (1980) 183}
\refdef\rPc\Pc{V.\ts Pasquier, \NP {\bf B285} (1987) 162}
\refdef\rPf\Pf{V.\ts Pasquier, \JP {\bf A20} (1987) L1229}
\refdef\rPSa\PSa{V.\ts Pasquier and H.\ts Saleur, \NP {\bf B330}
(1990) 523}
\refdef\rRa\Ra{P.\ts Roche, `On the construction of integrable dilute
A-D-E models', 1992 ENSLAPP preprint}
\refdef\rSg\Sg{N.\ts Sochen, \NP {\bf B360} (1991) 613}
\refdef\rSh\Sh{R.\ts Steinberg, \TAMS {\bf 91} (1959) 493}
\refdef\rSi\Si{R.\ts Steinberg, \PJM {\bf 118} (1985) 587}
\refdef\rSBa\SBa{H.\ts Saleur and M.\ts Bauer, \NP {\bf B320} (1989)
591}
\refdef\rSZb\SZb{H.\ts Saleur and J.-B.\ts Zuber, `Integrable Lattice
Models and Quantum Groups', Proceedings of the 1990 Trieste Spring School
on String Theory and Quantum Gravity}
\refdef\rVa\Va{E.\ts Verlinde, \NP {\bf B300} (1988) 360}
\refdef\rZe\Ze{\AZm, {\it Sov. Sci. Rev., Physics}, {\bf v.2} (1980)}
%
%
\def\bar{\overline}
\def\hat{\widehat}
\def\({\left(}
\def\){\right)}
\def\[{\left[}
\def\]{\right]}

\def\Z{{\bf Z}}

\def\a{\alpha} \def\b{\beta} \def\g{\gamma} 
  \def\l{\lambda} \def\p{\phi}
\def\t{\theta}   
\def\G{\Gamma}

\def\B{\bullet}
\def\W{\circ}

\def\aB{a_{\{\B\}}}
\def\aW{a_{\{\W\}}}
\def\lB{l_{\{\B\}}}
\def\lW{l_{\{\W\}}}
\def\wB{w_{\{\B\}}}
\def\wW{w_{\{\W\}}}

\noblackbox
\Title{SPhT/92-053}{Partition Functions, Intertwiners and the Coxeter
Element}
\centerline{
Patrick Dorey\footnote{$^\dagger$}{dorey@poseidon.saclay.cea.fr}}
\bigskip\centerline{Service de Physique Th\'eorique de
Saclay\rlap,\foot{{\it Laboratoire de la Direction des Sciences
de la Mati\`ere du Commissariat \`a l'Energie Atomique}}}
\centerline{91191 Gif-sur-Yvette cedex, France}
\vskip .5in
The partition functions of Pasquier models on the cylinder, and the
associated intertwiners, are considered. It is shown that earlier
results due to Saleur and Bauer can be rephrased in a geometrical way,
reminiscent of formulae found in certain purely elastic scattering
theories. This establishes the positivity of these intertwiners in a
general way and elucidates connections between these objects and the
finite subgroups of $SU(2)$. It also offers the hope that analogous
geometrical structures might lie behind the so-far mysterious results
found by Di\ts Francesco and Zuber in their search for generalisations
of these models.
\Date{May 1992}

\newsec{Introduction}
Consider a critical lattice model defined, not on a torus, but on a
cylinder of circumference $l$ and length $l'$. Once boundary
conditions $A$ and $B$ have been specified at the two ends, a
partition function $Z_{AB}$ can be defined which, for large enough
lattices, will be a function of the ratio $l/l'$ alone. The
continuum limit being a conformal field theory, this function
should be expressible in terms of characters of a (perhaps
extended) Virasoro algebra --- just one copy in this case, since
the surface has a boundary. Thus an expression of the
form
\eqn\ZABdef{Z_{AB}(l,l')\sim\sum_{r}n^r_{AB}\chi_r(q),
\qquad q=e^{-\pi l/l'}}
is expected\ts\RF\rCl\Cl, where $r$ runs over the possible
algebra representations
appearing in the spectrum of $H_{AB}$, the Hamiltonian propagating
states (on a line length $l'$, with boundary conditions $A$ and $B$)
around the cylinder, and
$\chi_r(q)$ is the corresponding character. The numbers $n^r_{AB}$ are
thus multiplicities, and as such should be non-negative integers.
In theories where the $\chi_r$'s are the characters of a
(maximally-extended) chiral algebra, in terms of which the torus
partition function is diagonal, Cardy\ts\RF\rCm\Cm\ subsequently
indicated a
correspondence between the coefficients $n^r_{AB}$ and the fusion rules
\`a la Verlinde\ts\RF\rVa\Va\ of this algebra.

In certain specific cases for the Pasquier models -- generalised RSOS
models where the heights live on the Dynkin diagram of some simply-laced
Lie algebra $G$\ts\RF\rPc\Pc\ -- equation \ZABdef\ was
confirmed by explicit calculation in ref.\ts\RF\rSBa\SBa.
Among other things, the case where
the heights at the two ends are constrained to be equal to $a$ and $b$
respectively was examined
($a$ and $b$ labelling two nodes on the Dynkin diagram of
$G$), with the result
\eqn\Zijdef{Z^{(G)}_{ab}(l,l')
\sim\sum_{\l=1}^{h{-}1}V^{\l}_{ab}\chi_{1,\l}(q),}
where
\eqn\vdef{V^{\l}_{ab}=\sum_{s\in\{{{\rm exponents}\atop{\rm of~}G}\}}
{\p^{(s)}_{\l}\over\p^{(s)}_1}q^{(s)}_aq^{(s)}_b.}
In the first equation, $h$ is the Coxeter number of $G$ and
$\chi_{1,\l}$ is a Virasoro character from the first row of the Kac
table for the central charge $c=1-6{/}h(h-1)$ of the model, while in
the second $\p^{(s)}$ and $q^{(s)}$ are eigenvectors of the $A_{h-1}$
and $G$ Cartan matrices respectively, both with eigenvalue
$2{-}2\cos\pi s/h$. Explicitly, $\p^{(s)}_{\l}=\surd(2{/}h)\sin(\pi
s\l{/}h)$. If an exponent $s$ of $G$ occurs more than once, there is a
freedom to rotate among the corresponding eigenvectors $q^{(s)}$, but
this causes no ambiguity since their total contribution to \vdef\ is
a quadratic form.

The result \Zijdef\ actually holds even before the continuum limit is
taken, on replacing each $\chi_{1,\l}(q)$ on the right hand side
by $Z^{(A_{h{-}1})}_{1\l}(l,l')$, the partition function
for an $A_{h-1}$ model with boundary spins $1$ and $\l$, evaluated in
the {\it same} geometry as the left hand
side\ts\NRF\rDZa\DZa\NRF\rSg\Sg\refs{\rDZa,\rSg}. (Consistency with
\Zijdef\ follows from the fact that
$Z^{(A_{h{-}1})}_{1\l}(l,l')\sim\chi_{1,\l}(q)$ in the continuum
limit\ts\rSBa.) This result justifies the use of the term
`intertwiner' for the matrix $V^{\l}$, and connects with the more
algebraic notions
that for finite systems the Virasoro algebra should be replaced by
that of Temperley and Lieb\ts\RF\rPSa\PSa. Thus
for a lattice of width $l'$ with boundary conditions $a$ and $b$,
the space of states is spanned by the
set ${\cal P}^{(l')G}_{ab}$ of paths of length $l'$
running from $a$ to $b$ on the Dynkin diagram of $G$, and
this space supports a
representation ${\cal R}^{(l')G}_{ab}$ of the Temperley-Lieb algebra.
The (modified) trace for this representation decomposes as
\eqn\trdec{{\rm tr}_{{\cal R}^{(l')G}_{ab}}(.)=\sum_{\l}V^{\l}_{ab}
{\rm tr}_{{\cal R}^{(l')A_{h{-}1}}_{1\l}}(.)}
into traces over $A$-type representations
${\cal R}^{(l')A_{h{-}1}}_{1\l}$,
a formula which contains the above-mentioned
result for finite-geometry
partition functions as a special case\foot{Strictly speaking,
\trdec\ is not always a decomposition:
for $G=D_n$, and $a$ and $b$ two extremal nodes with only one of
them a spinor, there is only one non-zero term on the right hand side of
\trdec, at $\l=n{-}1$.}.
The $V^{\l}$'s are again multiplicities, but this time
of representations of Temperley-Lieb rather than of Virasoro.

In fact, the authors of \refs{\rDZa,\rSg}\ were primarily interested in
these questions as a warm-up exercise in their
search for generalisations of the Pasquier models.
This also led them study the numbers $V^{\l}_{ab}$ on a more axiomatic basis,
thinking of them as (non-negative integer valued)
matrix representations of a Verlinde algebra. The $V^{\l}$ defined
above exhaust the possibilities for the fusion algebras of
the affine $\widehat{SU(2)}_k$ Kac-Moody theories at level $k=h{-}2$.
For further information on all this material, see the review articles
\NRF\rDZb\DZb\NRF\rSZb\SZb\NRF\rDe\De\refs{\rDZb{--}\rDe}.

Despite the many results that were uncovered in the course of these
investigations, it is likely that there remain some underlying principles
yet to be discovered. To give a small example, the fact that the
the $V^{\l}_{ab}$ are non-negative, clear from their interpretation
as multiplicities, is not at all evident from \vdef\ or its
generalisations, and was only established case-by-case
(even their integrality is not immediately obvious).  For this reason it
seems worthwhile to understand the $\widehat{SU(2)}_k$
examples, which are at least
completely classified, as thoroughly as possible. The purpose of this
paper is to report an observation which may help towards this goal,
allowing \vdef\ to be rewritten in a geometrical way for which
integrality and non-negativity are manifest, and for which various other
apparent coincidences (such as a connection with the finite subgroups
of $SU(2)$) become rather less mysterious.
The necessary geometrical information is established in section two,
while following sections relate this to \vdef\ and discuss various
implications.

\newsec{Geometrical details}
To set up notations, this section starts with a brief review of
some relevant information about the Coxeter elements of the Weyl group
of a simply-laced Lie algebra $G$. For further explanations and
applications, see
\NRF\rSh\Sh\NRF\rKb\Kb\NRF\rBc\Bc\NRF\rCf\Cf\NRF\rDc\Dc\NRF\rDd\Dd%
\NRF\rCDa\CDa\NRF\rLWb\LWb\NRF\rFb{\Fb\semi\FLOa}%
\NRF\rBa\Ba\NRF\rFOa\FOa\refs{\rSh{--}\rFOa}.
This material will then be used to derive expressions for certain inner
products between roots and/or weights, some of which turn out to reproduce
\vdef. The restriction to the simply-laced cases avoids the need to
distinguish left and right eigenvectors of the Cartan matrix
$C_{ab}^{(G)}$, and will be sufficient for current needs.

Splitting the simple roots of $G$ into two internally orthogonal sets,
$\Delta=\{\a_{\B}\}\cup\{\a_{\W}\}$,
according to a two-colouring of the Dynkin diagram
of $G$, $\sum_{\B'}$, $\sum_{\W'}$, $\sum_s$ will denote summation
over the black indices, the white indices and the exponents of $G$,
respectively. The eigenvectors $q^{(s)}$ of $C^{(G)}_{ab}$
satisfy
\eqn\qdef{C^{(G)}q^{(s)}=(2-2\cos\t_s)q^{(s)}}
where $\t_s=\pi s/h$ and $s$ is one of the exponents. These
eigenvectors will be normalised to length one, and the residual phases
picked so as to satisfy
\eqn\qprop{q^{(s)}_{\B}=q^{(h-s)}_{\B}\qquad
q^{(s)}_{\W}=-q^{(h-s)}_{\W}.}
If $s=h/2$ is an exponent, then this choice in fact fixes which roots are
black and which white -- for all indices in the white set, $q^{(h/2)}_{\W}$
vanishes. While the choice \qprop\ serves to fix various quantities
which arise during the working, the final results will not depend on
it.

Letting $r_i$ denote the Weyl reflection for the simple root $\a_i$,
set
\eqn\wwdef{\wB=\prod_{\B'}r_{\B'}
\qquad \wW=\prod_{\W'}r_{\W'}}
so that $w=\wB\wW$, the product of two involutions,
is a Coxeter element. (The utility of this particular choice of
Coxeter element was pointed out by Steinberg\ts\RF\rSh\Sh.)
An invariant subspace for $\wB$, $\wW$ and hence for $w$ is spanned by
\eqn\adef{\aB^{(s)}=\sum_{\B'}q^{(s)}_{\B'}\a_{\B'}
\qquad \aW^{(s)}=\sum_{\W'}q^{(s)}_{\W'}\a_{\W'}.}
The invariance stems from the facts, not too hard to verify, that $\wB$
and $\wW$ act on these vectors in the following way:
\eqnn\aprops
$$\eqalign{
\wB\aB^{(s)}=-\aB^{(s)}
 &\qquad\wB\aW^{(s)}=\aW^{(s)}+2\cos\t_s\aB^{(s)}\cr
\wW\aW^{(s)}=-\aW^{(s)}
 &\qquad\wW\aB^{(s)}=\aB^{(s)}+2\cos\t_s\aW^{(s)}\cr}\eqno\aprops$$
Following from these relations,
$|\aB^{(s)}|=|\aW^{(s)}|=1$ (except if $h/2$ is an exponent, in which
case $|\aB^{(h/2)}|=\surd 2$ and $\aW^{(h/2)}=0$), and
$(\aB^{(s)},\aW^{(s)})=-\cos\t_s$.
Dual to \adef, the simple roots can be expanded in terms of the invariant
subspaces:
\eqn\adual{\a_{\B}=\sum_sq^{(s)}_{\B}\aB^{(s)}
\qquad\a_{\W}=\sum_sq^{(s)}_{\W}\aW^{(s)}.}
The duality referred to here is between the set of points on the Dynkin
diagram and the set of exponents; there is of course also the vector
space duality between the simple roots $\a_a$ and the fundamental weights
$\l_b$, defined through $(\a_a,\l_b)=\delta_{ab}$.
The invariant subspaces for $w$ are
also spanned by objects constructed from these weights, namely
\eqn\ldef{\lB^{(s)}=\sum_{\B'}q^{(s)}_{\B'}\l_{\B'}
\qquad \lW^{(s)}=\sum_{\W'}q^{(s)}_{\W'}\l_{\W'},}
in terms of which the fundamental weights themselves expand as
\eqn\ldual{\l_{\B}=\sum_sq^{(s)}_{\B}\lB^{(s)}
\qquad\l_{\W}=\sum_sq^{(s)}_{\W}\lW^{(s)}.}
To check the various properties, note that
$(\lB^{(s)},\aB^{(s)})=(\lW^{(s)},\aW^{(s)})=1/2$ (or $1$, $0$
respectively if $s=h/2$), and that as a result of \qprop,
\eqn\alprop{\eqalign{
\aB^{(s)}=\aB^{(h-s)}&\qquad\aW^{(s)}=-\aW^{(h-s)}\cr
\lB^{(s)}=\lB^{(h-s)}&\qquad\lW^{(s)}=-\lW^{(h-s)}\cr}}
In addition, $|\lB^{(s)}|=|\lW^{(s)}|=1/(2\sin\t_s)$ (or
$1/\surd 2$ and $0$ if $s=h/2$), and the angle between $\lB^{(s)}$ and
$\lW^{(s)}$ is $\t_s$. Rather than write down yet more identities, the
data is summarised in figure 1. Noting from \aprops\ that, in the
subspace depicted, $\wB$ acts as the reflection leaving $\lW^{(s)}$ fixed
and vice
versa, it is then clear that $w$ itself is a rotation (in the
direction from $\lB$ to $\lW$) of $2\t_s=2\pi s/h$.

To see this action explicitly, a mixed basis is perhaps the most
convenient, such as that provided by $\lB^{(s)}$ and $\aW^{(s)}$.
Examining figure 1, it can be seen that \ldual\ generalises to
\eqn\wldual{w^p\l_a=\sum_sq^{(s)}_a\(\cos(2p{+}u_a)\t_s\lB^{(s)}
+{\sin(2p{+}u_a)\t_s\over 2\sin\t_s}\aW^{(s)}\),}
a formula which holds good even if $h/2$ is among the exponents.
Both black and white indices have been accounted for here, via
the convention that $u_a=0$ if $a\in\{\B\}$, $u_a=1$ if
$a\in\{\W\}$.

Enough information has now been gathered to write down expressions for
various inner products. First, consider $(w^p\l_a,\l_b)$. From \wldual,
\eqn\ipi{\eqalign{(w^p\l_a,\l_b)=\sum_{s,s'}
q^{(s)}_a&\(\cos(2p{+}u_a)\t_s\lB^{(s)}
+{\sin(2p{+}u_a)\t_s\over 2\sin\t_s}\aW^{(s)}\)\cr
&\qquad\times q^{(s')}_b\(\cos u_b\t_{s'}\lB^{(s')}
+{\sin u_b\t_{s'}\over 2\sin\t_{s'}}\aW^{(s')}\).\cr}}
Now $(\lB^{(s)},\aW^{(s')})$ is zero for all $s,s'$, as are all other
terms if $s'$ is not equal to either $s$ or $h-s$. Noting for the
latter case that, no matter what the colour of the index $b$,
$q^{(s)}_b\cos u_b\t_{s}$ is unchanged under $s\rightarrow h{-}s$
while $q^{(s)}_b\sin u_b\t_{s}$ is negated, the
identities given above together with a little algebra reduce
\ipi\ to
\eqn\wlprod{(w^p\l_a,\l_b)=\sum_sq^{(s)}_aq^{(s)}_b
{\cos(2p+u_{ab})\t_s\over 2\sin^2\t_s},}
where $u_{ab}=u_a-u_b$.
This expression is almost \vdef, but not quite. For this,
the inner products between roots and weights are needed. Within the
set $\Phi$ of all the roots, a complete set of
orbit representatives for $w$ is provided by the elements
$\p_a=(1{-}w^{-1})\l_a$ \RF\rKb\Kb. For the Coxeter element in use here,
their relationship with the simple roots is
\eqn\pprop{\p_{\B}=\wW\a_{\B}\qquad\p_{\W}=\a_{\W}.}
A summary of some of the useful properties of these roots
in the context of purely elastic S-matrices
can be found in \RF\rDd\Dd, and here too they turn out to be a
convenient choice. The inner product of $\l_a$ with $w^{-p}\p_b$
follows very quickly from \wlprod, since
$$\eqalign{(\l_a,w^{-p}\p_b)&=(\l_a,w^{-p}(1-w^{-1})\l_b)\cr
 &=(w^p\l_a,\l_b)-(w^{p+1}\l_a,\l_b).\cr}$$
After a small amount of work,
\eqn\wlpprod{(\l_a,w^{-p}\p_b)=\sum_sq^{(s)}_aq^{(s)}_b
{\sin(2p+1+u_{ab})\t_s\over \sin\t_s}.}
The same procedure repeated one more time gives the final identity of
this section, namely
\eqn\wppprod{(\p_a,w^{-p}\p_b)=\sum_s2q^{(s)}_aq^{(s)}_b
\cos(2p+u_{ab})\t_s.}
The case $a{=}b$, $p{=}0$ gives a simple check that there have been no
mistakes in the working.

Equations \wlprod, \wlpprod\ and \wppprod\ allow various identities
between inner products to spotted, some of which were listed in
\refs{\rDc,\rDd}. Actually, the identities given in \rDc\ involved the
simple roots $\a_a$ rather than the orbit representatives $\p_a$; to
obtain formulae for their inner products from those given
above, start by substituting for $\p_a$ and/or $\p_b$ using \pprop.
Then results such as
$w^{-p}\wW=\wW w^p$, $\wB\l_{\W}=\l_{\W}$ and $\wW\a_{\W}=-\a_{\W}$,
together with the Weyl-group invariance of the inner product,
rapidly convert \wlpprod\ and \wppprod\ into the desired forms.

\newsec{Consequences}
Comparing \wlpprod\ with \vdef\ and recalling the specific form of the
$A_{h-1}$ eigenvectors gives observation advertised in the
introduction:
\eqn\obs{V^{2p+1+u_{ab}}_{ab}=(\l_a,w^{-p}\p_b).}
Since in \Zijdef\ the index $\l$ ran over {\it all} integers from $1$
to $h{-}1$, it might appear that only half of the $V^{\l}$'s have been
given a geometrical interpretation by \obs. However the other
$V^{\l}$'s are zero, a fact which can be traced to the existence of a
$\Z_2$ charge\ts\rDZa, the index colour. To see this it suffices to
note that, from \qprop\ and the form of $\p^{(s)}_\l$,
\eqn\termconj{
{\p^{(s)}_{\l}\over\p^{(s)}_1}q^{(s)}_aq^{(s)}_b=(-1)^{\l{-}1{-}u_{ab}}
{\p^{(h{-}s)}_{\l}\over\p^{(h{-}s)}_1}q^{(h{-}s)}_aq^{(h{-}s)}_b.}
Hence for the sum in \vdef\ to be non-vanishing, $\l{-}1{-}u_{ab}$ must
be even, $2p$ say, a requirement which exactly reproduces the cases
covered by \obs.

The root systems under discussion being simply laced, the fundamental
weights are dual to the simple roots and the integrality of the
$V^{\l}$'s is immediate. That they are non-negative is just a little harder,
but has in fact already been discussed, albeit in the apparently very
different context of exact S-matrices, in \refs{\rDd,\rFOa}.

Comparing \obs\ with \Zijdef, what is needed is a proof that the inner
products $(\l_a,w^{-p}\p_b)$ are non-negative for $p$ in the range
$1\le 2p{+}1{+}u_{ab}\le h{-}1$. This will certainly be true if
$w^{-p}\p_b\in\Phi^+$ for all such $p$, where $\Phi^+$ is the set of
positive roots. This property was relevant for a discussion of certain
analyticity properties of the S-matrix formulae given in \rDc, and
essentially algebraic demonstrations were indicated in
\refs{\rDd,\rFOa}; for the sake of variety the following gives a
rather more geometrical approach.

By the Perron-Frobenius theorem (and perhaps after an overall sign
change), $q^{(1)}_a>0$ for all $a$, $q^{(1)}$ being the eigenvector of
$C^G$ with the smallest eigenvalue. Hence the vector
\eqn\lidef{l^{(1)}=\lB^{(1)}+\lW^{(1)}=\sum_aq^{(1)}_a\l_a}
is a strictly positive linear combination of the fundamental weights.
As a result, for any $\a\in\Phi$,
\eqn\liprop{ \a\in\Phi^+\Longleftrightarrow (l^{(1)},\a)>0.}
Since $l^{(1)}$ lies in the $s{=}1$ eigenspace of $w$, this means that
the positivity of a root can be ascertained simply by looking at its
$s{=}1$ projection. Figure 2 illustrates this, showing the projections
of the roots in two typical orbits. (As each orbit contains nine
roots, this is in fact the $A_8$ case.) This gives a visual
characterisation of the positive and negative roots in each orbit,
and
it is now easy to see that if $h$ is even the positive roots are
$w^{-p}\p_b$ for $p=0,\dots, h/2-1$, while for $h$ odd ({\it ie} for
$A_{2n}$) the relevant range for $p$ is $0,\dots, (h{-}3)/2$ if
$b\in\{\B\}$, and $0,\dots, (h{-}1)/2$ if $b\in\{\W\}$. Since these
ranges always include those relevant for the $V^{\l}$ that appear in
\Zijdef, the non-negativity of these numbers
has now been shown to follow from general principles.

Less visually but more explicitly, the formulae in the last section
lead to the following expression:
\eqn\liwprod{ (l^{(1)},w^{-p}\p_b)=q^{(1)}_b{\sin\(2p+{3\over
2}-u_b\){\pi\over h}\over\sin{\pi\over 2h}}~.}
Positivity is thus equivalent to $0<2p+{3\over 2}-u_b<h$, or (adding
$u_a{-}1/2$ throughout) $2p+1+u_{ab}$ lying between $u_a-1/2$ and
$h-1/2+u_a$. Whatever the value of $u_a$ ($0$ or $1$), this
includes the range $1,\dots, h{-}1$ needed to establish that, as
claimed, the $V^{\l}$'s are non-negative.

To sum up, formula \Zijdef\ for the partition function can now be
rewritten as
\eqn\Zijdefi{Z^{(G)}_{ab}
\sim\sum_{w^{-p}\p_b\in\Phi^+}(\l_a,w^{-p}\p_b)
 \chi_{1,2p{+}1{+}u_{ab}}~.}
This expression appears to contain extra terms over \Zijdef,
since, while the range for $p$ implied by positivity of $w^{-p}\p_b$
includes that relevant for \Zijdef, it does not necessarily coincide
with it exactly. The possible extra terms involve $\chi_{1,0}$ or
$\chi_{1,h}$; however all is well since
their coefficients $(\l_a,w^{-p}\p_b)$ are
forced to be zero by the relation
\eqn\dplrel{ (\l_a,w^{-p}\p_b)=-(\l_a,w^{p{+}1+u_{ab}}\p_b)}
and the fact that if $2p{+}1{+}u_{ab}$ is equal to $0$ or $h$,
we also have $w^{-p}\p_b=w^{p{+}1+u_{ab}}$.

To give a little more geometrical sense to \Zijdefi, some notation can
be borrowed from \rDd. For any pair of roots $\a,\b\in\Phi$, define an
integer $u(\a,\b)$ modulo $2h$ by
\eqn\udef{u(w\a,\b)=u(\a,\b)+2,\qquad u(\a,\b)=-u(\b,\a),\qquad
u(\p_a,\p_b)=u_{ab}.}
Then $\pi u(\a,\b)/h$ is the signed angle between the $s{=}1$
projections of $\a$ and $\b$, and \Zijdefi\ becomes
\eqn\Zijdefii{Z^{(G)}_{ab}
\sim\sum_{\b\in\G^+_b}(\l_a,\b)
 \chi_{1,1{+}u(\p_a,\b)},}
where $\G_b^+$ is the intersection of $\G_b$, the $w$-orbit of $\p_b$,
with $\Phi^+$, the set of positive roots. That $u(\a,\b)$ has only
been defined modulo $2h$ causes no ambiguities since
$\chi_{r,s{+}2h}=\chi_{r,s}$ (recall that the characters involved here
are those for Virasoro central charge $1-6/h(h{-}1)$).

Suitably reinterpreted, the expressions \Zijdefi\ and \Zijdefii\ apply
equally to the expansions of partition functions in finite geometries,
and to the decompositions of the modified traces, equation \trdec. For
these two applications, the modulo $2h$ ambiguity in $u(\a,\b)$ should
be removed by imposing $0\le u(\a,\b)<2h$, since $\l=1+u(\p_a,\b)$
really should only run from $1$ to $h{-}1$.

To close this section, a remark on a curious coincidence. In
refs.~\refs{\rDc,\rDd}\ general formulae were given for the S-matrix
elements of the (simply-laced) affine Toda field theories, and of
certain perturbed conformal field theories. In either category, there
is a theory associated with each simply-laced Lie algebra $G$, having
$r={\rm rank}(G)$ particle types, one for each node on the Dynkin
diagram of $G$. The scattering amplitude for a pair of particles, of
types $a$ and $b$, is given by
\eqn\Sabdef{S_{ab}(\t)=\prod_{\b\in\G^+_b}\{1+u(\p_a,\b)\}^{(\l_a,\b)},}
a product of a number of functions $\{.\}(\t)$ of the rapidity $\t$.
The precise forms of these building blocks
differ between the affine Toda and perturbed conformal cases;
they can be found in \RF\rBCDSc\BCDSc.
(In fact, the formulae in \refs{\rDc,\rDd}\
involved products taken over the entire orbit $\G_b$ of certain
sub-blocks $\{.\}_{\pm}$, but it is easy to see that they are
equivalent to \Sabdef.) The formal similarity
between \Sabdef\ and \Zijdefii\ should be
clear, and gives a small practical application of the observation
\obs: should the numerical values of the coefficients $V^{\l}_{ab}$
ever happen to be needed, they can be read from the complete tables of
the affine Toda S-matrix elements given in \rBCDSc.
Whether this similarity is any more than a coincidence remains to be
seen.

\newsec{Connections with the McKay correspondence}
The McKay correspondence\ts\RF\rMj\Mj\ (see also
\NRF\rSi\Si\NRF\rKc{\Kc\semi\Kd}\refs{\rSi,\rKc})
is a bijection between the finite subgroups of $SU(2)$ and the affine
Dynkin diagrams of types $\hat A$, $\hat D$ and $\hat E$,
such that to each of the finitely-many irreducible representations
$\g_a$ of a given finite subgroup
$\G$ there is associated a node $a$ on the corresponding
affine Dynkin diagram $\hat G$. This association is encoded as
follows: if $\g$ is the two-dimensional representation of $\G$
provided by the original $SU(2)$, then
\eqn\mckdef{\g\otimes\g_a=\sum I^{(\hat G)}_{ab}\g_b~,}
where $I^{(\hat G)}$ is the incidence matrix of $\hat G$ (so $C^{(\hat
G)}=2{-}I^{(\hat G)}$). The trivial representation $\g_0$ is always
associated with the `extra' spot of $\hat G$,
corresponding to the negative of the
highest root for the non-affine algebra $G$.

The irreducible representations of $SU(2)$ itself are infinite in
number, there being one ($\pi_n$ say) for every $n=0,1,2,\dots,$ of
dimension $n{+}1$. On restriction these provide
representations $\pi_n|\G$
of the finite subgroups $\G$, which may now be reducible. In \rKc,
Kostant studied the decompositions
\eqn\kprob{\pi_n|\G=\sum_bm_n^b\g_b}
of these representations into $\G$-irreducibles.
In particular, for each irreducible representation $\g_b$ of $\G$
he computed the Poincar\'e series $P_{\G}(t)_b$,
encoding the multiplicities $m^b_n$ as
\eqn\pgdef{P_{\G}(t)_b=\sum_{n=0}^{\infty}m_n^bt^n,}
and found
\eqn\pgres{P_{\G}(t)_b={z(t)_b\over (1-t^A)(1-t^B)}~,}
where $A$ and $B$ are two $\G$-dependent integers constrained by
$A{+}B=h{+}2$, $AB=2|\G|$, and $z(t)_b$ is a polynomial in $t$,
of degree at most $h$. For the trivial representation $\g_0$,
$z(t)_0=1{+}t^h$, while for the remaining representations $\g_b$,
for which the index $b$ also identifies a spot on the non-affine diagram,
the expression for $z(t)_b$ invokes root system ideas very close to
those described in section 2. To give the explicit formula, a little
extra notation is needed. First, the two-colouring
$\Delta=\{\a_{\B}\}\cup\{\a_{\W}\}$ is alternatively labelled as
$\Delta=\Delta_1\cup\Delta_2$, with the requirement
that all the simple roots in
$\Delta_2$ should be orthogonal to $\psi$, the highest root of $G$. As
is clear from the form of the affine diagrams, this can be arranged for
all root systems except $A_{\rm even}$, a case which Kostant explicitly
excluded. Correspondingly, $\wB$ and $\wW$ are rewritten as
$w_1$ and $w_2$ (not
necessarily respectively), the requirement on $\Delta_2$ implying that
$w_2\psi=\psi$. For $n\in\Z_+$, let $w_n=w_1$ if $n$ is odd and
$w_n=w_2$ if $n$ is even ($w_n$ should not be confused with the simple
Weyl reflections, denoted $r_a$ above), and finally set
$w^{[n]}=w_nw_{n-1}\dots w_1$, $w^{[-n]}=w_1w_2\dots
w_n=(w^{[n]})^{-1}$. Then for $b\neq 0$,
\eqn\zres{z(t)_b=
\sum_{n=1}^{h-1}(\l_b,w^{[n-1]}\psi-w^{[n]}\psi)\,t^n.}
(Kostant also gave various other forms for this expression,
but \zres\ is the most relevant here.)

Since the expression \zres\ uses machinery similar to that employed
in earlier sections, the observation of Di\ts Francesco and
Zuber\ts\ref\rDZobs{P.\ts Di\ts Francesco and J.-B.\ts Zuber,
unpublished, reported in \rDe.} that the coefficients of the
$z(t)_b$ were to be found as certain of the $V^{\l}_{ab}$ should
not now be too surprising. The remainder of this section will show
precisely how this works, drawing on various ideas from \rKc\ to put
\zres\ into a form closer to \obs.

To start, rewrite the coefficient of $t^n$ in \zres\ as
\eqn\newcoeff{z_{b,n}=(\psi,w^{[-(n{-}1)]}(\l_b-w_n\l_b)).}
Now $w_n\l_b$ is equal to $\l_b-\a_b$ or $\l_b$, depending on whether
the $b$ is associated with the same subset ($\Delta_1$ or $\Delta_2$)
as $w_n$ or not. Thus roughly half of the coefficients $z_{b,n}$ are
zero. To avoid overburdening the notation, assume for the time being
that $\{\a_{\B}\}=\Delta_2$, $\{\a_{\W}\}=\Delta_1$ (this may conflict
with \qprop, but recall that that particular choice only
affected intermediate stages of the working and not the final results).
Thus $\a_b\in\Delta_{2-u_b}$, and $\l_b-w_n\l_b$ is non-zero only
if $n$ modulo $2$ is equal to $2{-}u_b$, that is if $n=2p{+}2{-}u_b$
for some $p$; in such cases it is equal to $\a_b$. Referring back to
\pprop, it is straightforwardly checked that, whatever the colour of
$\a_b$,
\eqn\pcheck{w^{[-(2p{+}1{-}u_b)]}\a_b=w^{-p}\p_b,}
and so equation \zres\ becomes
\eqn\zresi{z(t)_b=\sum_{1\le 2p+2-u_b\le h-1}
(\psi,w^{-p}\a_b)\,t^{2p{+}2{-}u_b}.}
Now for the $\hat D$ and $\hat E$ affine diagrams, the negative of the
highest root joins to the remaining, non-affine part of the diagram by
just a single link, connecting it to the simple root $\a_f$ say. Hence
$(\psi,\a_a)=\delta_{af}$, and so $\psi=\l_f$. For $\hat A_{{\rm odd}}$,
the remaining case, $\psi$ has inner product $1$ with both extremal
roots on the non-affine diagram ($\a_f$ and $\a_{\bar f}$ say), and so
is equal to $\l_f+\l_{\bar f}$. In either case, the specification of
$\Delta_2$ means that $\a_f$ and/or $\a_{\bar f}$ belong to
$\Delta_1=\{\a_{\W}\}$, so $u_f{=}u_{\bar f}{=}1$ and $2p{+}2{-}u_b$
can be replaced by $2p{+}1{+}u_{fb}$ or $2p{+}1{+}u_{\bar fb}$.
Substituting all of this into \zresi, comparing
with \obs\ and recalling that the $V^{\l}$'s not accounted for by
\obs\ are automatically zero establishes that
\eqn\zderes{z_{b,n}=V^n_{fb}}
for $D$ and $E$, while
\eqn\zares{z_{b,n}=V^n_{fb}+V^n_{\bar fb}}
for $A_{{\rm odd}}$. Equations \zderes\ and \zares\ exactly reproduce
the observations of \rDZobs. (Interestingly, \zares\ also holds for
$A_{{\rm even}}$.) The choice to set
$\{\a_{\B}\}=\Delta_2$, $\{\a_{\W}\}=\Delta_1$ clearly should have
no bearing on these final results, and indeed it is not too hard to check
explicitly that \obs\ is unchanged if the black and white roots are
swapped -- the only points to note are that such a swap negates
$u_{ab}$, sends $w$ to its inverse {\it and} changes the
definition \pprop\ of each $\p_a$.

Given the correspondence between \Sabdef\ and \Zijdefii, the considerations
of this section also apply to the exact S-matrices of affine Toda type;
this was (very briefly) mentioned as a `note added' in ref.\ts\rDd.

\newsec{Conclusions}

These conclusions fall naturally into two parts: first, questions that
remain in the $\widehat{SU(2)}$ case; and second, the (potentially
more interesting) question of generalising the above constructions to
other models associated with Lie algebras of higher rank.

To start the discussion for $\widehat{SU(2)}$, it is worth recalling
that the Pasquier models in their continuum limits provide
representatives for many of the unitary $c<1$ conformal field
theories classified by Cappelli, Itzykson and Zuber\ts\RF\rCIZa\CIZa,
but not all of them.
For central charge $c=1-1{/}h(h{-}1)$, the possible
modular invariant partition functions
are labelled by a pairs $(G,G')$ of Lie
algebras, with Coxeter numbers $h{-}1$ and $h$ respectively -- this
forces one of $G,G'$ to be of type $A$. The torus partition function
of the Pasquier model associated with the algebra $G$ (of Coxeter number
$h$) corresponds to the pair $(A_{h-2},G)$ in the continuum limit, and
so theories labelled by the pair $(G_{h-1},A_{h-1})$
are missed.  However,
lattice models have now been found which are expected to yield
the $(G,A)$ partition functions\ts\RF\rRa\Ra. It would be interesting
to generalise the calculations of \rSBa\ to cover these models, and to
find out whether the expansions corresponding to \Zijdef\ also hide
geometrical features similar to those outlined above.

Partition functions on the torus, and the associated
issues of modular invariance\ts\RF\rCk\Ck, may seem rather
disconnected from the discussions above of partition functions on
surfaces with boundaries. There are at least two reasons why
this is not so. As already mentioned in the introduction, Cardy\ts\rCm\
has established that in certain cases the link between boundary
conditions, fusion rules and modular invariance is rather close.
Unfortunately, this requires knowledge of expansions analogous to
\Zijdef\ for a complete set of boundary states invariant under the
maximally-extended chiral algebra of the theory, which is generally
larger than just Virasoro. Furthermore, the expansions should be in
characters of this larger algebra. These conditions are not met by the
expansions \Zijdef\ beyond the (rather trivial) $A$ case, so it is not
possible to apply Cardy's arguments directly here. Of more
immediate relevance is the second point,
an empirical observation made in \rDZb\ that
certain of the $V^{\l}$'s encode the decomposition of the extended
conformal blocks for many modular invariant partition functions,
thereby probing a finer structure than that revealed by examining
characters of the maximal algebra alone. To be a little more precise,
attention should first be restricted to the so-called `type I'
theories\ts\rDZa, that is theories for which the toroidal partition
function is diagonal, a sum of squared moduli (the same restriction
applied to the discussion in \rCm, in fact). For the $(A,G)$ or
$(G,A)$ $c<1$ theories, or for the $\widehat{SU(2)}_k$ affine Kac-Moody
models labelled by a single algebra $G_k$, type I partition
functions are found for $G=A$, $D_{\rm even}$, $E_6$ and $E_8$.
In these cases, a particular subset $T$ of the nodes of the Dynkin
diagram of $G$ is chosen (these subsets are listed in \rDe), and in
addition a special node $a_0$ is picked, such that $q^{(1)}_{a_0}$
is minimal (alternatively put, $a_0$ labels the lightest particle in
the corresponding affine Toda theory). Then the partition function on
the torus is simply
\eqn\partdef{Z=\sum_{b\in T}Z_b,}
where (in obvious notation)
\eqna\pdf
$$ Z_b^{(A,G)}={1\over 2}
\sum_{r=1}^{h-2}|\sum_{\l=1}^{h-1}V^{\l}_{a_0b}\chi_{r,\l}|^2;\qquad
Z_b^{(G,A)}={1\over 2}
\sum_{s=1}^{h-1}|\sum_{\l=1}^{h-2}V^{\l}_{a_0b}\chi_{\l,s}|^2
\eqno\pdf a$$
for the $c<1$ models, while
$$Z_b^{(G)}= |\sum_{\l=1}^{h-1}V^{\l}_{a_0b}\chi_{\l}|^2\eqno\pdf b$$
for an affine $\widehat{SU(2)}$ modular invariant, with $\chi_{\l}$ an
affine rather than Virasoro character in this case. Combining these
expressions with the formula \obs\ for the $V^{\l}$'s shows that the
geometry of root systems certainly has a r\^ole to play in the
construction of the $ADE$ modular invariants, though there are clearly
many elements of this which are obscure. In particular, it would be
interesting to find a geometrical interpretation of the type~I /
type~II distinction, and for the special subset $T$ of simple roots
referred to above. An understanding of the modifications
necessary to Cardy's arguments to cope with the expansions
\Zijdef\ might be a help in this regard, as might a direct derivation
of the Virasoro decomposition of the Pasquier model toroidal partition
functions from the lattice models, analogous to that achieved by
Saleur and Bauer\ts\rSBa\ on the cylinder\foot{Note, the treatment given
by Pasquier in \RF\rPf\Pf\ is rather more indirect than this,
in that he first
relates the toroidal partition functions to sums of partition
functions of certain other models, the so-called $f$-models.}.
Nevertheless,
even as it stands the observation does shed
a little light on some of the $ADE$ numerology that has been observed
among the modular invariants. For example, in \RF\rIa\Ia\ it was
noted that the conformal blocks for the $E_8$ modular invariants
can be read from two of the Poincar\'e polynomials of the binary
icosahedral subgroup of $SU(2)$. In the case of $E_8$, the node $a_0$
relevant for the modular invariants is the same as the node $f$
which arose in the discussion of the McKay correspondence in the
previous section, and so the observation is consistent with equations
\zderes\ and \pdf{}. Note that this particular coincidence, between the
Poincar\'e polynomials of a
finite subgroup of $SU(2)$ and a modular invariant, does not
generalise beyond $E_8$, consistent with the fact that it is only for
this case that $a_0=f$.

The above have been rather specific questions. More generally, it
would be good to have a better understanding of {\it why} there should
be such a geometric interpretation for the quantities $V^{\l}$.
For the purely elastic S-matrices, the general expression
\Sabdef\ seems less mysterious once it is seen how naturally such
formulae solve the bootstrap equations. Given the formula \Zijdefii\ for
the partition functions, an immediate thought is that there might be
analogues of the bootstrap equations relating partition functions on
the cylinder with different boundary conditions. However the variable
$q$ seems a poor candidate to replace the rapidity $\t$ in \Sabdef;
much more natural would be to re-introduce the spectral parameter.
This in turn is reminiscent of the close ties that
exist between factorisable S-matrices and lattice
models\ts\RF\rZe\Ze, ties which may ultimately explain the formal
similarities between equations \Sabdef\ and \Zijdefii. Nevertheless,
and despite various promising signs, a physically-motivated set of
equations for the partition functions \Zijdef, or some small
generalisation of them, has proved elusive. It may be that, just
as the purely elastic S-matrices are too simple to exhibit any
Yang-Baxter structure, so the bootstrap structure is absent for the
Pasquier models, and would only be seen in some larger class of objects,
within which both the purely elastic S-matrices and the Pasquier models
would be found as degenerate special cases.

Finally to the question of generalisations beyond the $c<1$ Pasquier
models, to intertwiners associated with algebras of higher rank than
$SU(2)$. The approach adopted in \refs{\rDZa,\rSg}\ was to search for
algebraic and graph-theoretic features of the Pasquier models and
their intertwiners, and then to place these in a wider context. More
general graphs than the simply-laced Dynkin diagrams arose, but
nevertheless formulae exactly analogous to \vdef\ were found, for
which many of the features described above (in particular
non-negativity) continued to hold. The main point of this paper has
been that, to understand the intertwiners in the $\widehat{SU(2)}$
case, it is necessary to add some geometrical insight to the algebra
and graph theory. It would be very interesting if
geometrical structures could be found lying behind the many mysterious
results found in \refs{\rDZa,\rSg} for the higher-rank algebras,
and indeed to see what these
structures might be. Since the case of $\widehat{SU(2)}$ has already
exhausted all finite reflection groups, the search will have to be
quite wide, and might perhaps lead to something genuinely new.

\bigskip\noindent{\bf Acknowledgements}\smallskip\nobreak
I would like to thank M.\ts Bauer, P.\ts Di\ts Francesco, C.\ts
Itzykson, V.\ts Pasquier, N.\ts Sochen and J.-B.\ts Zuber
for many helpful discussions and explanations of their work.
I am grateful to the European Community for a
grant under the EC Science Programme.
\listrefs
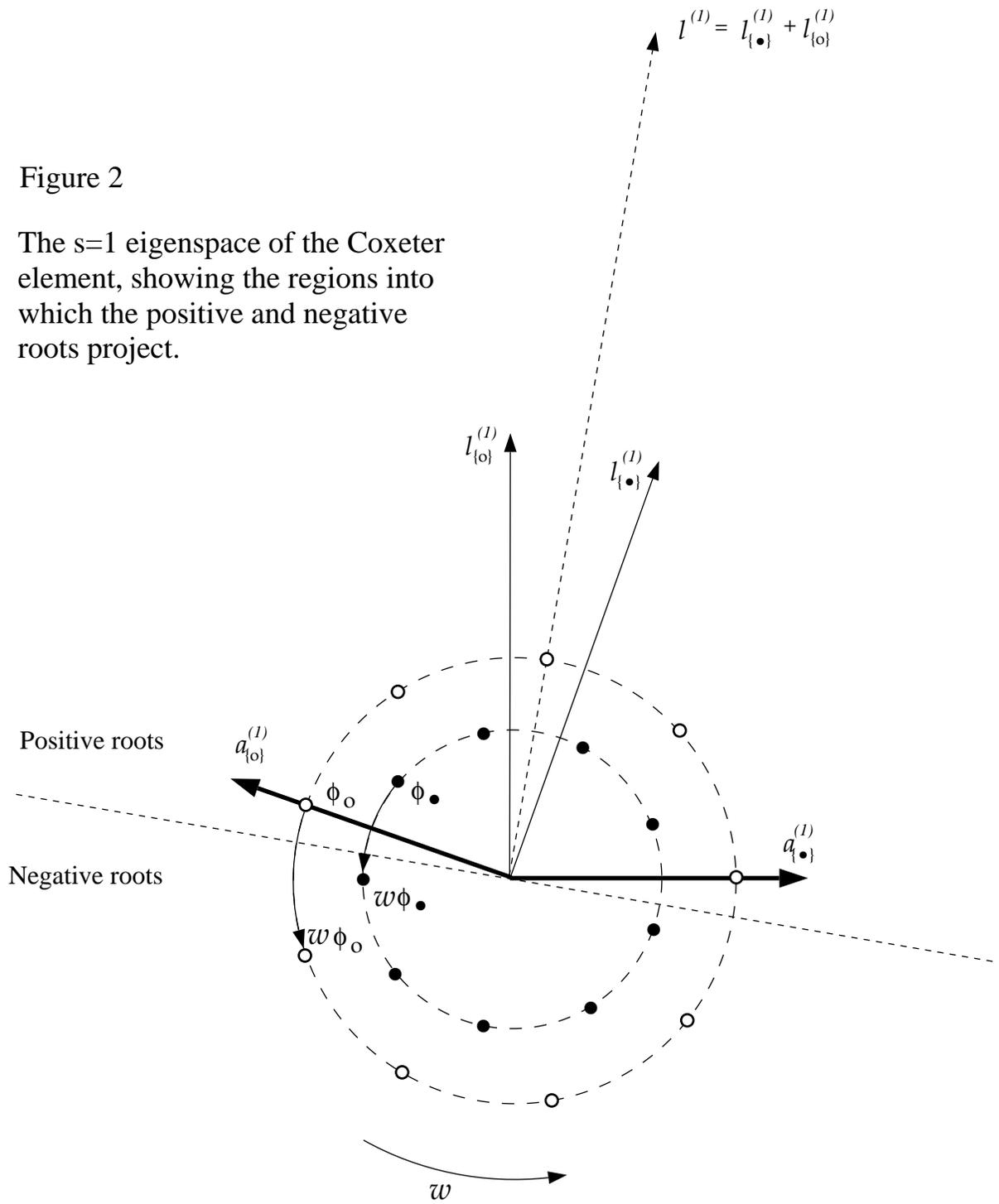
\bye